\documentstyle[prd,aps,epsf,epsfig]{revtex} 
\begin{document}
\draft
\twocolumn[\hsize\textwidth\columnwidth\hsize\csname@twocolumnfalse\endcsname
%
\title{Collective dynamics of two-mode stochastic oscillators}
\author
{A. Nikitin$^1$, Z. N\'eda$^{2,*}$ and T. Vicsek$^3$}
\address{
$^1$Department of Physics, Saratov State University, 410026 Saratov, Russia\\
$^2$Department of Physics, University of Notre Dame, IN 46656, USA\\
$^3$Department of Biological Physics, E\"otv\"os Lor\'and University, Budapest, Hungary}
\maketitle
\centerline{\small (Last revised \today)}

\begin{abstract}
We study a system of two-mode stochastic oscillators coupled through their collective
output. As a function of a relevant parameter four qualitatively 
distinct regimes of collective behavior are observed.
In an extended region of the
parameter space the periodicity of the collective output is enhanced by the 
considered coupling. This system can be
used as a new model to describe synchronization-like
phenomena in systems of units with two or more oscillation modes. 
The model can also explain how periodic 
dynamics can be generated by coupling largely stochastic units.
Similar systems could be responsible for the emergence of rhythmic
behavior in complex biological or sociological systems. 

\end{abstract}

\vspace{2pc}
]

\vspace{1cm}


Collective dynamics of stochastic systems shows a great variety 
of interesting phenomena like pulsating patterns \cite{hempel}, spiral waves
\cite{spiral},
synchronization \cite{synchron} or stochastic resonance 
\cite{sr}. While the collective dynamics
of single mode stochastic oscillators is relatively well understood
\cite{single}, 
the dynamics of stochastic oscillators with many possible modes is less studied 
\cite{old}, and leads to
interesting and new effects. An example in this sense 
is our recent
study on the peculiar dynamics of the rhythmic applause \cite{nature}, where
the fascinating dynamics appears as a result   
of two different frequency clapping modes \cite{PRE}. Many other 
biological or physical systems exist, which perform stochastic oscillations
with different modes. As examples we mention the unicellular alga Gonyaulax
polyedra where circadian oscillators with two different periods 
have been shown to
co-exist \cite{morse}, the thalamocortical relay neurons which can generate
either spindle (7-14 Hz) or delta (0.5-4 Hz) oscillations \cite{wang}, and the
hippocampal CA3 model which spontaneously can generate four different
rhythms \cite{tateno}. 
When the system can switch between the available modes 
as a function of some threshold condition, new and interesting collective
behavior is observed.

The present study reports on such results for an
ensemble of coupled two-mode stochastic oscillators. 
As a function of a threshold condition we reveal different type of
synchronized and unsynchronized phases. 
We also show that rhythmic collective behavior can be generated
by suitably coupling largely stochastic units, i.e. the periodicity of 
the total output relative to the output of one stochastic oscillator can be
strongly enhanced. As an immediate application we offer a new
and realistic description for the dynamics of the rhythmic applause, 
reproducing all experimentally observed characteristics.

Our system is composed by identical pulse-coupled two-mode 
stochastic oscillators. Their cycle can be performed in two modes: 
$A\rightarrow B_I \rightarrow C \rightarrow A$ or $A\rightarrow B_{II} 
\rightarrow C \rightarrow A$, respectively. The periods corresponding to these
two modes $T_I$ and $T_{II}$ are given as
$T_{I}=\tau_A + \tau_{B_I} + \tau_{C}$ and $T_{II}=\tau_A + \tau_{B_{II}} + 
\tau_{C}$, where
$\tau_A$, $\tau_B$ and $\tau_C$ are time intervals spent in state $A$, $B$
and $C$, respectively. The stochastic part of the dynamics is {\em state A},
and $\tau_A$ is a stochastic variable with distribution:
\begin{equation}
P(\tau_A)=  \frac{1}{\tau^*} exp(-\frac{\tau_A}{\tau^*}),
\end{equation}
($\tau^*=\langle \tau_A \rangle$). State $A$ should be imagined and 
modeled with an escape dynamics of a stochastic field-driven particle from
a potential valley of depth $U$. If the stochastic force-field is totally 
uncorrelated with $\langle\xi \rangle=0$ and $\langle \xi(t) \xi(t+\tau)\rangle=D \delta(\tau)$ we
get a distribution of escape times given in (1) with:
$\tau^* \sim exp(U/D)$.
In analogy with the well-known FitzHugh-Nagumo system \cite{FHN},
state $A$ corresponds to a stochastic reaction time of the neuron fire.
This causes all the experimentally observed fluctuations in the rhythmic human 
activities \cite{BiPh,clapflu}. 
In state $B$ and $C$ the dynamics is deterministic, and
corresponds to the relaxation of the neurons \cite{BiPh,science}. 
{\em State B} represents
a "waiting time" or the rhythm giving part of the cycle. In biological systems
this is a period the individual units want to impose, and usually this is the 
longest part of the cycle. The length of state B ($\tau_{B_I}$ or
$\tau_{B_{II}}$) distinguishes between the two modes. We have chosen
$\tau_{B_{II}}=2\tau_{B_I}$. The output of the units 
is in {\em state C}. During this state the oscillator emits a constant
intensity pulse of strength $1/N$ where $N$ is the number of
oscillators in the system. The output of the whole system at a given 
moment is                                               
\begin{equation}                                     
f=\sum_{i=1}^{N} f_i,
\end{equation}
where $f_i$ is the output of oscillator $i$, $f_i=1/N$ if the given 
oscillator is in state $C$ and $f_i=0$ otherwise. This total output is
the origin of the coupling and shifts the oscillators between
their operating modes. The rules for the evolution of the
system are as follows: (i) oscillators start with randomly selected modes
and phases and follow the stated dynamics; (ii) there is a fixed output intensity, $f^*$,
for the system; (iii) after completing the dynamics in state $A$,
each oscillator will choose to
operate either in mode I or mode II; (iv) if at that moment $f<f^*$ the oscillator
will operate in mode I, otherwise will follow mode II.
The above dynamics has the tendency to keep the average total output 
as close as possible to $f^*$. 
Since each oscillator has a fixed  output intensity, this can be achieved 
only through switching
between the available modes. In this sense the proposed rules are natural, 
making our model realistic.

The evolution of a single uncoupled unit is simple since it
preserves the starting mode ($s$). The average value of its period is:
\begin{equation}
\nonumber \langle T \rangle=\langle \tau_A + \tau_{B_s} + \tau_C \rangle= 
\tau^* + \tau_{B_s} + \tau_C. 
\end{equation} 
Its relative standard deviation
\begin{equation}
R=\frac{\sqrt{\langle T^2 \rangle- \langle T \rangle^2}}{\langle T \rangle}=
\frac{\tau^*}{\tau^* + \tau_{B_s} + \tau_C},
\end{equation}
shows an increasing tendency as a function of $\tau^*$.

In contrast with the simple uncoupled case, the dynamics of the coupled units is
complex. Generally the oscillators will shift irregularly between their two 
modes. As a function of $f^*$ many interesting 
regimes are observed. 

For the limiting values $f^*=0$ or $1$ the dynamics is trivial. For $f^*=1$
we always have $f<f^*$ and all units operate in mode I. The $f$ output
randomly fluctuates due to both the initial random phases and the 
stochastic nature of $\tau_A$. As the number of units increases, the variance 
of this fluctuation decreases.
The average value of the collective output 
$\langle f(f^*=1) \rangle=\langle f \rangle_1$,
is a function of 
$\tau_C$ and $\langle T_I \rangle $ only:
\begin{equation}
\langle f \rangle_1=\frac{\tau_C}{\langle T_I \rangle}.
\end{equation}
The above considerations will apply in the
$N>>1$ limit for the $f^*> \langle f \rangle_1$ cases as well.
In this manner we can
analytically predict that for high enough $f^*$ values the units operate as 
simple uncoupled stochastic oscillators in mode I. For $f^*\le 0$
all units operate in mode II, and again there is no effective coupling between the
units. The total signal will randomly fluctuate, and 
due to the larger $\langle T_{II} \rangle $ value its mean will be
smaller than the
one measured in the $f^*> \langle f \rangle_1$ cases:
\begin{equation}
\langle f \rangle_0=\langle f(f^*=0) \rangle 
=\frac{\tau_C}{\langle T_{II} \rangle} < \langle f \rangle_1 
\end{equation}
For $0<f^*< \langle f \rangle_1$ we have
a nontrivial regime, where the coupling is effective
and the oscillators switch between their modes. This region
will be studied by computer simulations. We can numerically 
follow both the dynamics of a selected unit and the total $f$  
output of the system. For fixed $\tau_{B_I}$, $\tau_{B_{II}}$ and $\tau_C$
values the parameters governing the system dynamics are $f^*$ and $\tau^*$.
We choose $\tau_{B_I}=0.4$ and 
$\tau_C=0.1$. By mapping the relevant $\{f^*,\tau^*\}$ parameter-space, four different
regimes (phases) can be revealed. {\bf Phase I}, is an unsynchronized regime
with $\langle f \rangle= \langle f 
\rangle_0$ (Fig.~1a). {\bf Phase II}, is a synchronized regime with quasi-periodic total output
and large oscillation periods close to the $\langle T_{II} \rangle $ value 
(Fig.~1b). {\bf Phase III}, is a synchronized regime with quasi-periodic output and
slow oscillation periods close to the $\langle T_I \rangle$ value (Fig.~1c).
{\bf Phase IV}, is an unsynchronized regime with $\langle f \rangle= \langle f 
\rangle_1$ (Fig.~1d).

Changing the values of $\tau_B$ and $\tau_C$ leads to qualitatively 
similar phases in the parameter space.
From the phase-space  sketched in Fig.2 we learn that by increasing the value of
$f^*$ the systems collective output changes from the unsynchronized 
phase I behavior to the synchronized phase II and phase III dynamics, leading
finally to 
the unsynchronized phase IV regime. As $\tau^*$ is increasing phase III 
disappears ($\tau^*\ge0.5$) and the $f^*$ interval where synchronization
is present (phase II and III) diminishes. In agreement with our previous 
analytic
justifications, in phase I we have $\langle f \rangle=\langle f \rangle_0$ 
(Fig.~1a), 
in phase IV, $\langle f \rangle = \langle f \rangle_1$ and for $f^*> 
\langle f \rangle_1$ 
the collective behavior of the
system always corresponds to the phase IV one. 

In the $0<f^*< \langle f \rangle_1$ regime the dynamics of a single unit is non-trivial.
The units stochastically shift between mode I and mode II
oscillations. In this regime although the movement of each unit is largely
stochastic, their collective behavior leads to a periodic output. 
In order to characterize numerically
the enhancement in the periodicity we define a measure for it.
Let us denote the output signal as a function of time as $f(t)$. We
can define an error function, $\Delta(T)$, which characterizes numerically 
how strongly the 
$f(t)$ signal differs from a periodic signal with period $T$
\begin{equation}
\Delta(T)=\frac{1}{2M} lim_{x\rightarrow \infty} \frac{1}{x} \int_0^x \mid f(t)-f(t+T)
\mid dt, 
\end{equation}
where
\begin{eqnarray}
\nonumber M=lim_{x\rightarrow \infty}\frac{1}{x} \int_0^x \mid f(t)-<f(t)> \mid dt \\
<f(t)>=lim_{x\rightarrow \infty} \frac{1}{x} \int_0^x f(t) dt 
\end{eqnarray}
The general shape of the $\Delta(T)$ curve as a function of $T$ is sketched on
Fig.~3. For any $f(t)$ oscillating function we have an initially increasing tendency
at small $T$ values, after which for  $T=T_{m}$ a minimum
($\Delta_m$) is reached. One can state that $T_m$ is the best approximation for
the $f(t)$ signals period, and the "periodicity level" of the signal is
characterized by:
\begin{equation}
p=\frac{1}{\Delta_m}
\end{equation}  
We can compute this parameter both for one oscillator working independently
($p_1$) in the long period mode (where the effect of stochasticity on the period
is smaller) and for the whole system ($p$). The ratio $p/p_1$ will
characterize the enhancement in the periodicity. 
We have analyzed throughout the $\{\tau^*,f^*\}$ parameter space this enhancement in
the periodicity. Results for $N=200$
oscillators are given in Fig.~4. From this data it is
clear that there is a quite large parameter space (corresponding to small values
of $f^*$ and large limits for $\tau^*$) where the enhancement in the
periodicity is considerable. The computed best periods 
($T_m$) are in agreement with the phases previously 
discussed and sketched in
Fig.~2.  Increasing the number of coupled oscillators will further enhance the
periodicity of the total output. As an example results for $\tau^*=0.4$ and
$f^*=0.1$, are presented in Fig.~5.
It is also interesting to note the stochastic resonance type effect \cite{sr} 
for the enhancement in the periodicity (Fig. 4). 
For fixed $f^*$ values there is an
optimal noise level ($\tau^*$) in the oscillators dynamics, where a maximal
$p/p_1$ enhancement ratio is achieved through the considered coupling.

As an immediate application one can create a new model and
explanation for the phenomenon of rhythmic applause \cite{PRE}.  
Different phases of the collective output (Fig.2) reproduces
the regimes observed in collective clapping.  The high intensity (phase IV)
unsynchronized phase corresponds to the initial thunderous 
clapping of the spectators following an exceptional performance. Phase II
describes the rhythmic applause where 
spectators clap in unison with a relatively long period. Phase III with 
a shorter period and partly synchronized output is characteristic for the 
unstable transition intervals from synchronized to unsynchronized 
clapping. The low intensity unsynchronized phase I dynamics is characteristic
to the clapping of a non-enthusiastic audience. Modeling the clapping
individuals by the proposed two-mode stochastic oscillators, 
and the coupling between
them by the proposed threshold criteria 
is realistic and in agreement with experimental results: i) the
clapping sound is pulse like and is reproduced by state C of the oscillators;
ii) the clapping period of one individual has a 
fluctuating nature \cite{clapflu}
and is modeled by state A of the dynamics; iii) the rhythm of the clapping 
is modeled by state B; iv) our previously reported measurements  
on clapping individuals \cite{nature} revealed the existence of the 
two distinct clapping modes with short and longer periods, respectively; 
v) the period of the longer clapping mode 
was found do be roughly two times larger than the one of the
short clapping mode. In view of this new model the characteristic interplay
between synchronized and unsynchronized regimes in the rhythmic applause 
should be a consequence of a peculiar dynamics in the $f^*$ threshold, and
should have a psychological origin. It is clear that after a bad performance
the $f^*$ threshold is low and leads to phase I type collective response.
For an enthusiastic audience $f^*$ is big and high intensity 
collective response forms, resembling the total output 
of the units in phase IV. By lowering the
level of $f^*$ (fatigue or just resting...?) synchronized collective response
arises (phase III) which corresponds to the rhythmic applause.

In our opinion the presented model has possibilities for understanding
and modeling the origin of other rhythmic biological or sociological phenomena
\cite{examples}. The most interesting aspect of our results is 
the finding that regular periodic 
output can be obtained by coupling a large number of stochastic units.
Similar phenomena should be responsible for the emergence of rhythmic behavior
in complex biological or sociological systems. 

The present study was done during a common fellowship at Collegium Budapest-
Institute of Advanced Study (Hungary). We acknowledge the professionally
motivating academic atmosphere from the Collegium.











\begin{figure}[htp]
\epsfig{figure=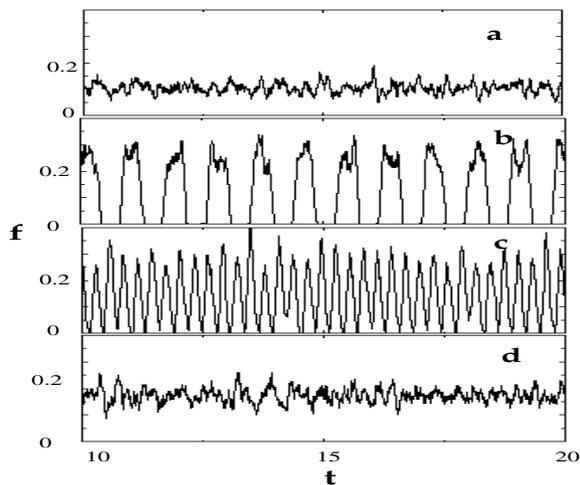,height=2.5 in,width=3.0in,angle=-0}
\caption{Total $f$ output of 200 coupled two-mode oscillators 
($\tau^*=0.1$, $\tau_{B_I}=0.4$, $\tau_C=0.1$ and $f^*=-0.1$ (a), $f^*=0.1$ (b), 
$f^*=0.2$ (c), $f^*=0.3$ (d)).}
\label{fig1}
\end{figure}


\begin{figure}[htp]
\epsfig{figure=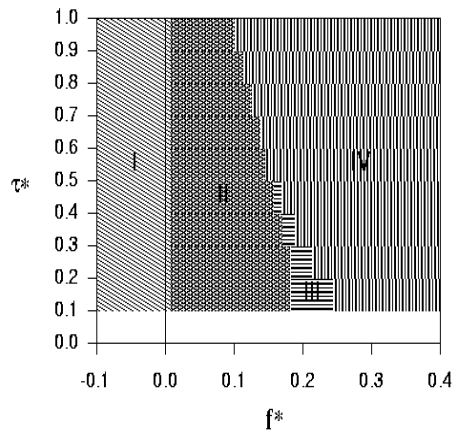,height=2.5in,width=2.5in,angle=-0}
\caption{Phases in the $\{f^*,\tau*\}$ parameter-space ($N=200$,
$\tau_{B_I}=0.4$, $\tau_C=0.1$).
}
\label{fig2}
\end{figure}


\begin{figure}[htp]
\epsfig{figure=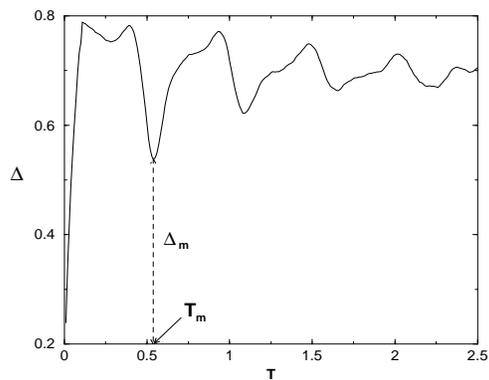,height=2.0in,width=2.5in,angle=-0}
\caption{Characteristic shape of $\Delta(T)$. Determining the
$\Delta_m$ and $T_m$ parameters. ($\tau*=0.1$, $f^*=0.3$, 
$\tau_{B_I}=0.4$, $\tau_C=0.1$, $N=200$)
}
\label{fig3}
\end{figure}


\begin{figure}[htp]
\epsfig{figure=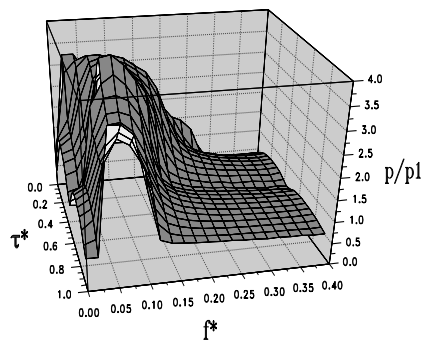,height=2.3in,width=3.0in,angle=-0}
\caption{Enhancement in the periodicity ($p/p_1$) as a function of the 
$\tau^*$ and $f^*$ relevant parameters. (Results for $N=200$ oscillators,
$\tau_{B_I}=0.4$, $\tau_C=0.1$)
}
\label{fig4}
\end{figure}


\begin{figure}[htp]
\epsfig{figure=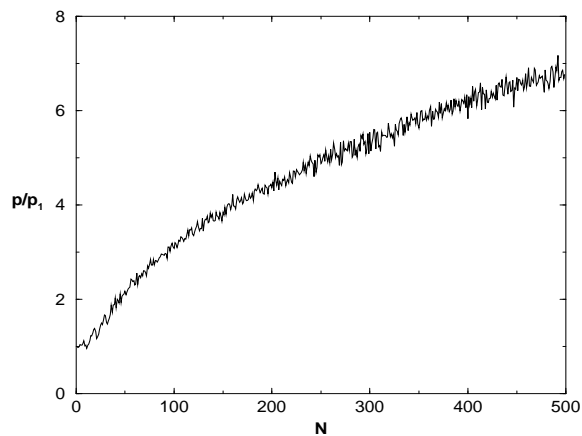,height=2.3in,width=3.0in,angle=-0}
\caption{Enhancement in the periodicity of the output ($p/p_1$) as a function
of the number of oscillators in the system. ($f^*=0.1$ and $\tau^*=0.4$
$\tau_{B_I}=0.4$, $\tau_C=0.1$)
}
\label{fig5}
\end{figure}
\end{document}